\documentclass[pre,twocolumn]{revtex4}
\usepackage[english]{babel}
\usepackage{amsmath}
\usepackage{amssymb}
\usepackage{epsfig}

\begin{document}
\title{Systems of vortices in a binary core-shell Bose-Einstein condensate}
\author{Victor P. Ruban}
\email{ruban@itp.ac.ru}
\affiliation{Landau Institute for Theoretical Physics RAS,
Chernogolovka, Moscow region, 142432 Russia}

\date{\today}

\begin{abstract}
A trapped Bose--Einstein-condensed mixture of two types of
cold atoms with significantly different masses
has been simulated numerically within the coupled
Gross--Pitaevskii equations. A configuration consisting
of a vortex-free core and a shell penetrated by quantum
vortices is possible in the phase separation regime.
The dynamic properties of vortices in the shell are 
determined by several parameters. Physically implementable 
parametric domains corresponding to long-lived strongly
nonstationary systems of several vortices
attached to the core have been sought. A number of
realistic numerical examples of three vortex pairs existing
for many hundreds of characteristic times have been presented.

\vspace{4mm}

DOI: 10.1134/S0021364022601579
\end{abstract}

\maketitle

\subsection*{Introduction}

In the physics of ultracold Bose--Einstein-condensed gases,
a great amount of interest is focused on
multicomponent mixtures consisting of different
chemical (alkaline) elements, or different isotopes of a
single element, or identical isotopes in different
(hyperfine) quantum states. These systems in static
and dynamic properties are much richer than single-component 
Bose--Einstein condensates \cite{mix1,mix2,mix3,mix4,mix5}.
Binary Bose--Einstein condensates are experimentally
implemented, for example, in the
${}^{87}$Rb-${}^{87}$Rb \cite{Rb87-Rb87-tun},
${}^{85}$Rb-${}^{87}$Rb \cite{mix-Rb85-87,Rb85-Rb87-tun},
${}^{39}$K-${}^{87}$Rb  \cite{K39-Rb87-tun},
${}^{41}$K-${}^{87}$Rb  \cite{K41-Rb87-tun-1,K41-Rb87-tun-2}, 
${}^{23}$Na-${}^{39}$K  \cite{Na23-K39-tun}, and
${}^{23}$Na-${}^{87}$Rb \cite{Na23-Rb87-tun} systems.

Sufficiently dilute mixtures of Bose gases with
atomic masses $m_j$ in the zero temperature limit can be
theoretically studied within the coupled Gross--Pitaevskii 
equations describing the evolution of the corresponding wavefunctions:
\begin{equation}
i\hbar\partial_t\Psi_j=-\frac{\hbar^2}{2m_j}\nabla^2\Psi_j+
\Big[V_j({\bf r})+\sum_k G_{jk}|\Psi_k|^2\Big]\Psi_j.
\label{GPE-coupled}
\end{equation}
Here, $V_j({\bf r})$ is the external potential acting on the $j$th 
component of the mixture and $G_{jk}$ is the symmetric
matrix of nonlinear interactions given by the expression \cite{mix2}
\begin{equation}
G_{jk}=2\pi \hbar^2 a_{jk}(m_j^{-1}+m_k^{-1}),
\end{equation}
where $a_{jk}$ are the scattering lengths. The Gross--Pitaevskii equations
provide the simplest but very comprehensive model of coupled superfluid systems
(cf. the review of studies of helium \cite{SV1987}). In particular,
since Eq. (\ref{GPE-coupled}) does not include cross terms in the
kinetic energy, the Andreev--Bashkin effect is completely 
absent; i.e., the superfluid velocity of one component 
does not contribute to the current of the other component \cite{AB1975,V2022}.

Some coefficients $G_{jk}$ usually depend on the back-
ground uniform magnetic field and can be varied in a
wide desired range using Feshbach resonances \cite{RMP2010}. 
A sufficiently strong cross repulsion between two types
of matter waves switches on the spatial separation of
phases under the condition \cite{separation,AC1998}
\begin{equation}
G_{12}^2-G_{11}G_{22}>0.
\end{equation}
A relatively narrow domain wall between phases is
characterized by the effective surface tension \cite{mix4,tension}.
The spatial separation is mainly responsible for many
interesting phenomena such as bubble dynamics \cite{bubbles},
quantum analogs of classical hydrodynamic instabilities 
(Kelvin--Helmholtz \cite{KHI1,KHI2}, Rayleigh--Taylor \cite{RTI1,RTI2,RTI3},
and Plateau--Rayleigh \cite{capillary}), parametric
instability of capillary waves at the interface \cite{param_inst-1,param_inst-2},
complex textures in rotating binary condensates 
\cite{mix-sheet-1,mix-sheet-2,topo_defects}, 
three-dimensional topological structures 
\cite{vortex-mol,wall-annih-1,wall-annih-2,vortex-wall,handles}, 
and capillary flotation of dense droplets in
trapped immiscible Bose--Einstein condensates \cite{R2021-2}.

It is also important that trap potentials are generally
different for different types of atoms. For example, the
potential of an optical trap formed by laser radiation has the form
\begin{equation}
V_j({\bf r})=-\frac{\alpha_j}{2}I({\bf r}) + m_j gz,
\end{equation}
where ${\alpha_j}$ is the polarizability of an atom of the $j$th
type, $I({\bf r})$ is the time-averaged square of the electric
field of laser radiation, and $g$ is the gravitational acceleration. 
Since the dependence of the polarizability on
the frequency of the electromagnetic wave is specific
to each component, selecting the frequency of the
optical field, one can theoretically control the relative strength
of the trap and the spatial positions of minima of the
potential for each component \cite{polariz1,polariz2,polariz3}. It is implied
below that either the polarizability ratio is $\alpha_2/\alpha_1=m_2/m_1$
(at which the gravitational force does
not ``expand'' the minima of the potentials in space),
or the system is in free fall state, where it is possible to
vary the important parameter
$$
\alpha= \alpha_2 m_1/(\alpha_1 m_2).
$$
In this work, the mentioned possible variations of the
system parameters are taken into account in the
numerical search for optimal regimes for the long-term 
dynamics of vortex excitations in spherical
``core-shell'' structures. The ground state of binary
Bose--Einstein condensates sometimes has such configurations 
\cite{gr1,gr2,gr3,shell2022}. A core consisting of one
component is formed near a sufficiently deep and/or wide
minimum of the external potential; the shell of the
other component surrounds the core.

\begin{figure}
\begin{center}
\epsfig{file=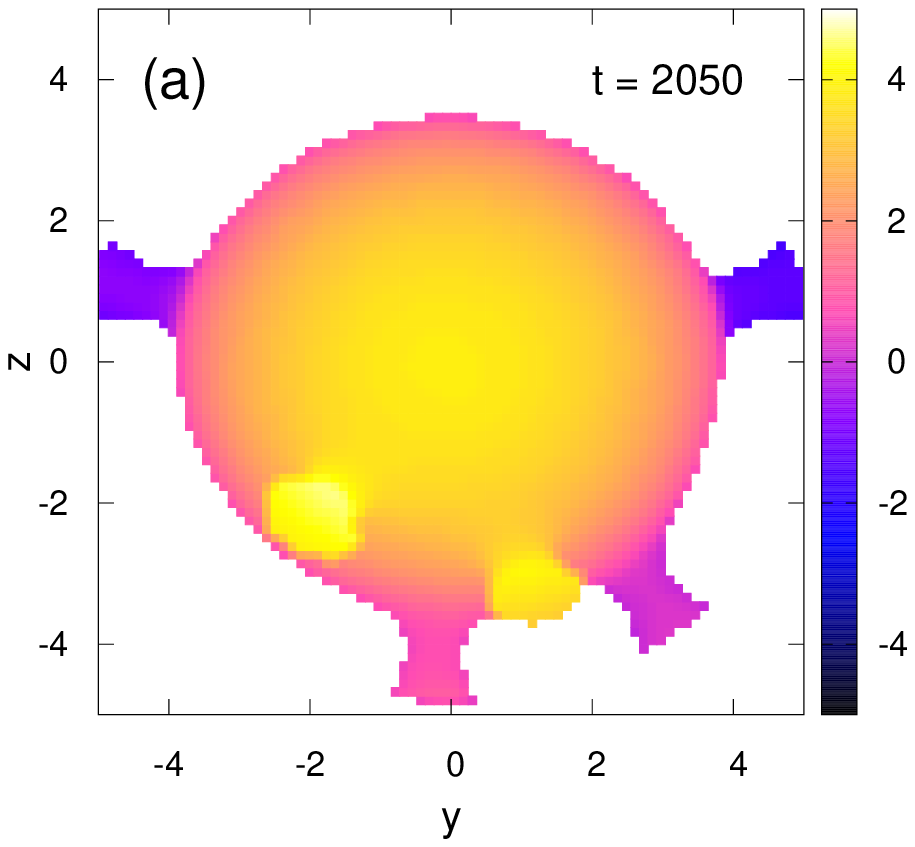, width=75mm}\\
\epsfig{file=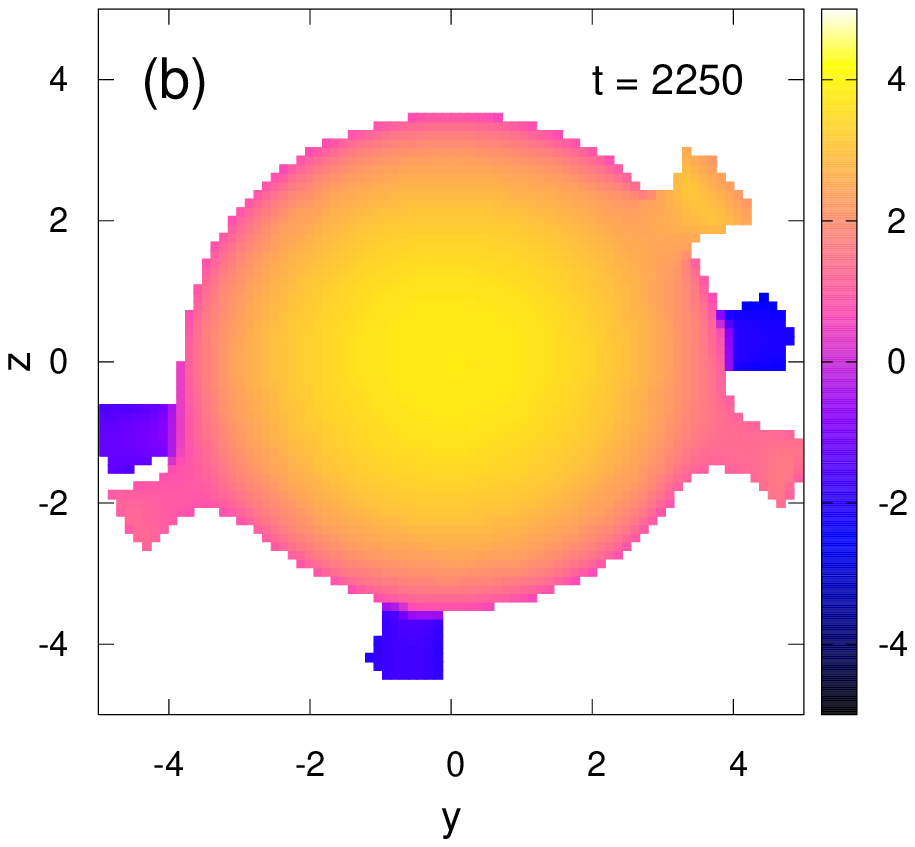, width=75mm}\\
\epsfig{file=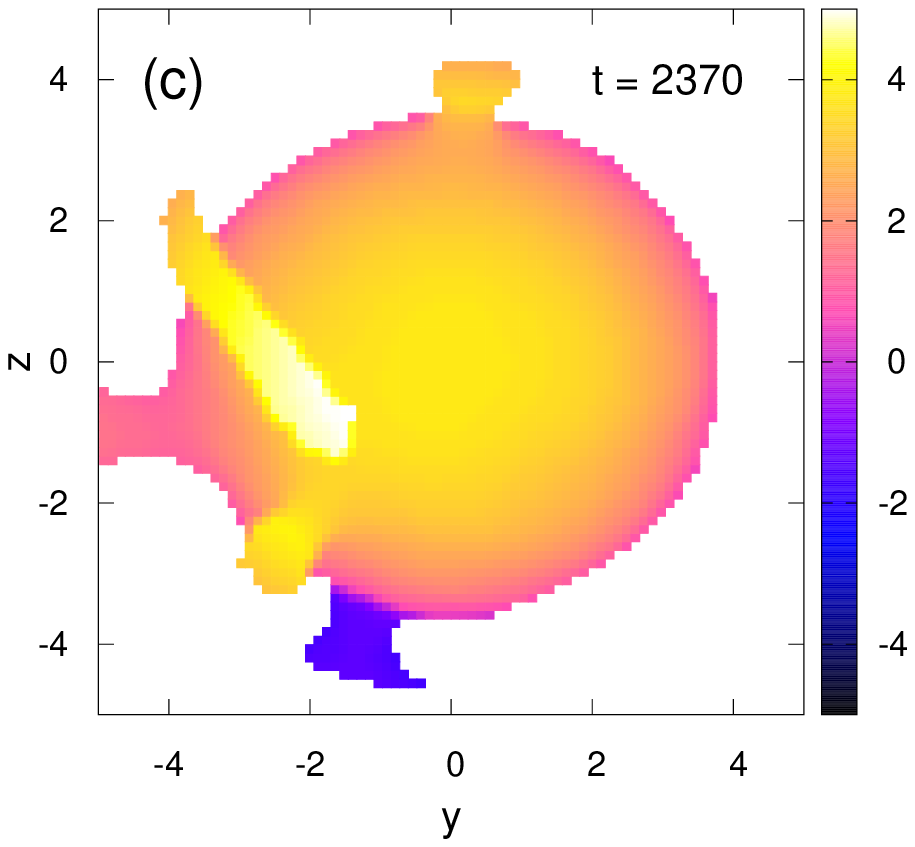, width=75mm}
\end{center}
\caption{Example of a long-lived nonstationary 
system of three pairs of attached vortices in the
quasi-two-dimensional regime at the compensated gravitational
force (i.e., at $\alpha=1$ ). The color scale corresponds to the $x$
coordinate of the points of the numerical lattice that are
the closest to the surface specified by the equation
$|A({\bf r},t)|^2 = 0.5|A_0({\bf r})|^2$. Only points inside the region
$|A_0({\bf r})|^2 > 0.2\mu_1$ are shown; for this reason, the outer ends
of vortex filaments are ``truncated''.
}
\label{example-quasi-2D} 
\end{figure}

Below, we consider the dynamics of several quantized 
vortex filaments each penetrating the shell
toward the core or away from the core (negative or
positive vortex, respectively) \cite{R2021_attached_v}. An example of the
nontrivial dynamics of such attached vortices is shown
in Fig. 1 and in video \cite{video1}. Similar topological
structures in the ${}^3$He-${}^4$He mixture (a ${}^4$He droplet
immersed in liquid ${}^3$He) were qualitatively considered
in \cite{V2000}. The total charge of vortices is always zero. The
core remains vortex-free. Two regimes are possible
depending on the ratio of the shell thickness to the
core size. A more simple (quasi)two-dimensional
regime similar to the motion of concentrated vortices
in atmospheres of some planets occurs in thin shells.
The dynamics in this case is approximately described
by the classical model of point vortices on a curved
surface \cite{R2017,shell-BEC-1,shell-BEC-2}. 
The core in this case serves as an
almost immobile potential step supporting the shell
from inside. The other, three-dimensional, regime is
much more complex \cite{R2021_attached_v} and occurs when the shell
thickness is comparable with the core radius and the
core is significantly involved in the dynamics of the
system. Depending on the parameters, the effect of
the core can be both favorable for the long-term storage 
of vortices despite their intense interaction and unfavorable.

As recently shown in \cite{R2021_attached_v}  for the case of equal
atomic masses, there is a region of parameters where
the undesired process of formation of a close vortex-antivortex 
pair with the subsequent separation of a
vortex filament from the interface with the core is suppressed 
because of the optimal choice of the coefficients of 
the nonlinear interaction. In this work, such
favorable conditions are determined for the first time
in numerous numerical experiments for some mixtures
with significantly different atomic masses. It is
interesting that configurations where the core consists
of lighter atoms and the shell is formed of heavier
atoms are optimal for the three-dimensional dynamic
regime. In this case, the parameter $\alpha$ equal to the ratio
of squares of the eigenfrequencies of the harmonic
trap is not unity.

\subsection*{Basic parameters of the model}

The quadratic approximation is used for external potentials. 
Let a harmonic trap be characterized by transverse frequencies 
$\omega_1=\omega$ and $\omega_2=\sqrt{\alpha}\omega$
for atoms of the first type with the mass $m_1$ forming the
shell and for atoms of the second type with the mass $m_2$
forming the core, respectively. The time, length, and
energy will be represented in units of
$\tau=1/\omega$, $l_{\rm tr}=\sqrt{\hbar/\omega m_1}$, 
and $\varepsilon=\hbar\omega$, respectively. In real experiments
at the frequency $\omega/2\pi\sim$100 Hz, the characteristic length
of the trap $l_{\rm tr}$ is about several microns.
Since the typical scattering lengths are $a_{jk}\sim a\sim 100 a_0$, 
where $a_0$ is the Bohr radius, i.e., several nanometers, 
the inequality $(l_{\rm tr}/a)\sim 10^3\gg 1$ necessary 
for the applicability of the Gross--Pitaevskii equations is certainly satisfied.

The equations of motion for the complex wavefunctions $A({\bf r},t)$ (shell)
and $B({\bf r},t)$ (core) can be written in the dimensionless form
\begin{eqnarray}
i\dot A&=&-\frac{1}{2}\nabla^2 A+\left[V+|A|^2+g_{12}|B|^2\right]A,
\label{GP1}\\
i\dot B&=&-\frac{1}{2m}\nabla^2 B+
\left[m\alpha V+g_{21}|A|^2+g_{22}|B|^2\right]B,
\label{GP2}
\end{eqnarray}
where $m=m_2/m_1$ is the ratio of atomic masses, 
$g_{jk}=G_{jk}/G_{11}$ are the orthonormalized 
nonlinear coefficients, and $V=(x^2+y^2+\lambda^2 z^2)/2$
is the dimensionless potential including the anisotropy
parameter $\lambda$, taken as $\lambda=1.1$ to emphasize that the
strict spherical symmetry is not necessary.

In this normalization, the conserved numbers of
trapped atoms are given by the formulas
\begin{eqnarray}
&&N_1=\frac{l_{\rm tr}}{4\pi a_{11}}\int |A|^2 d^3{\bf r}=(l_{\rm tr}/a_{11}) n_1,\\
&&N_2=\frac{l_{\rm tr}}{4\pi a_{11}}\int |B|^2 d^3{\bf r}=(l_{\rm tr}/a_{11}) n_2.
\end{eqnarray} 
According to these formulas, the realistic numbers $N_1,N_2\sim 10^6$
correspond to $n_1,n_2\sim 10^3$.

As known, equilibrium states are characterized by
two chemical potentials $\mu_1$ and $\mu_2$. In our case, 
$\mu_1\gg 1$ and $\mu_2\gg 1$ so that the background profiles of the
particle number densities in the shell and core are given in
the Thomas--Fermi approximation
\begin{eqnarray}
&&|A_0|^2\approx [\mu_1-V(x,y,z)],
\label{AA_eq}\\
&&|B_0|^2\approx [\mu_2-m\alpha V(x,y,z)]/g_{22},
\label{BB_eq}
\end{eqnarray} 
respectively. Thus, the effective transverse size of the
condensate is $R_\perp=\sqrt{2\mu_1}$, whereas the characteristic
thickness of vortex filaments in the shell (in the three-dimensional regime) 
is estimated as $\xi\sim 1/\sqrt{\mu_1}$. Since
the chemical potentials in the systems considered
below are about several tens (the value $\mu_1=18$ was
used in numerical experiments), the physical thickness 
of vortices $\xi l_{\rm tr}$ is still larger than the scattering
length by orders of magnitude, as should be the case.

In this approximation, the equilibrium shape of the
core surface is determined by the requirement of the
approximate equality of ``hydrodynamic pressures''
$P_1=|A_0|^4/2$ and $P_2=g_{22}|B_0|^4/2$
(disregarding the surface tension $\sigma\sim |A_0|^3$), i.e.,
\begin{equation}
[\mu_1-V(x,y,z)]=[\mu_2-m\alpha V(x,y,z)]/\sqrt{g_{22}}.
\end{equation}
For the convective stability, the ``stratification parameter'' should exceed unity:
\begin{equation}
S=m\alpha/\sqrt{g_{22}}>1.
\end{equation}
At finite $\mu_1$ and $\mu_2$ values, this inequality should be
satisfied with some margin because the surface tension
against the inhomogeneous background has a destabilizing effect. 
The optimal $S$ values for the three-dimensional 
dynamics of attached vortices are in the range of 1.2--1.5, 
whereas the hard stratification with large $S$ values is better for 
the quasi-two-dimensional regime.

\begin{figure}
\begin{center}
\epsfig{file=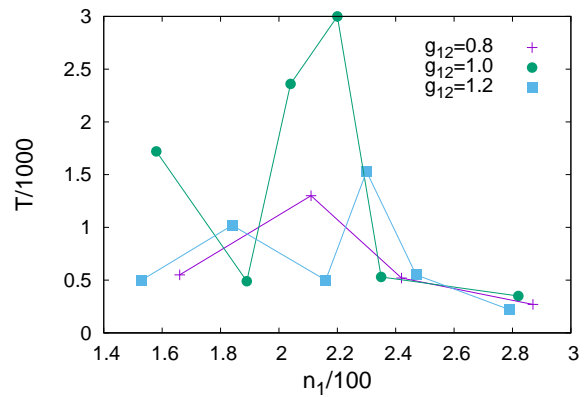, width=80mm}
\end{center}
\caption{Characteristic lifetime of the system
of three pairs of vortices versus the number of atoms of the
shell in the ${}^{23}$Na-${}^{87}$Rb mixture for three $g_{12}$ values.
The dependences are irregular because averaging over the initial 
configurations of vortices was not performed.
}
\label{quasi-2D-times} 
\end{figure}

\begin{figure}
\begin{center}
\epsfig{file=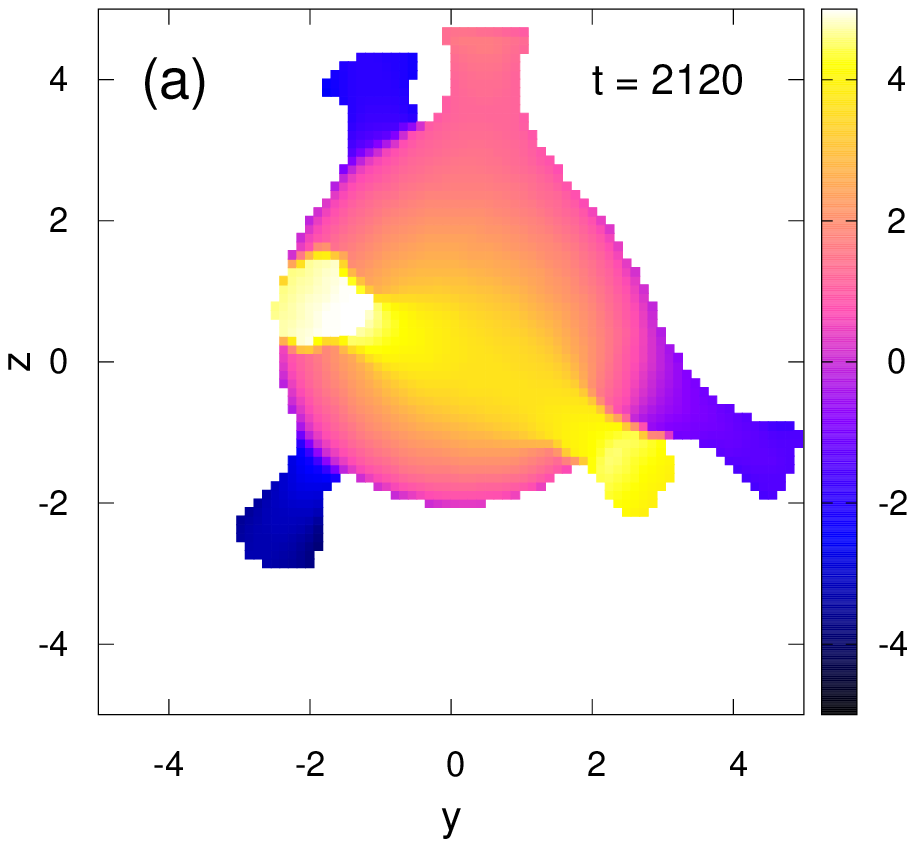, width=75mm}\\
\epsfig{file=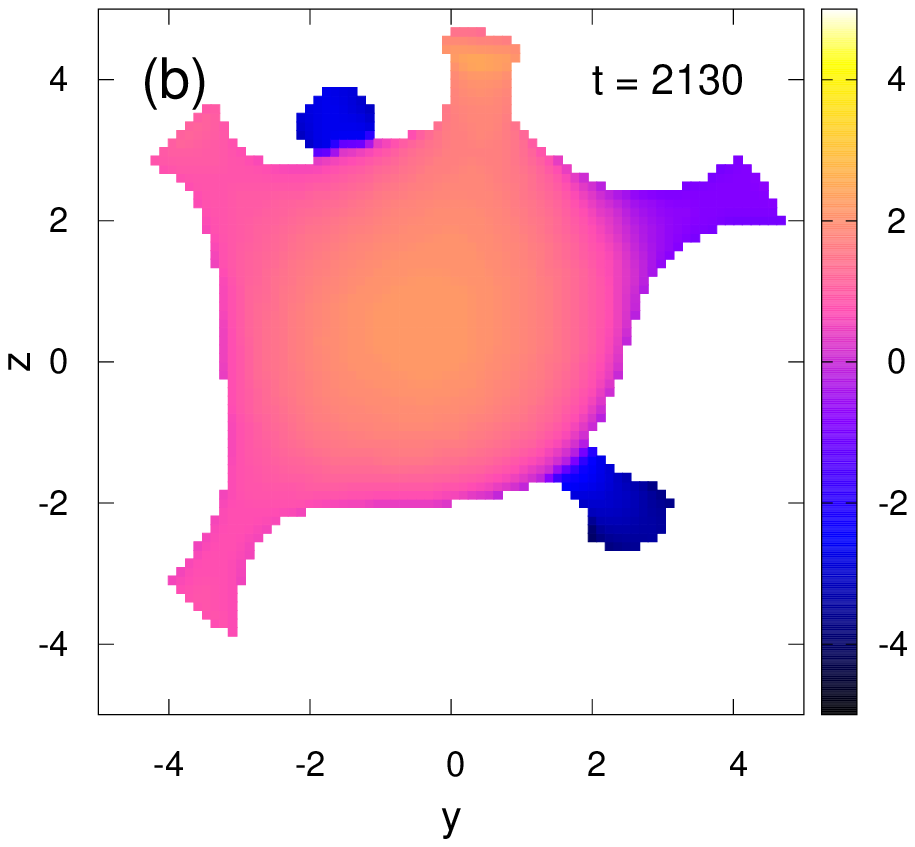, width=75mm}\\
\epsfig{file=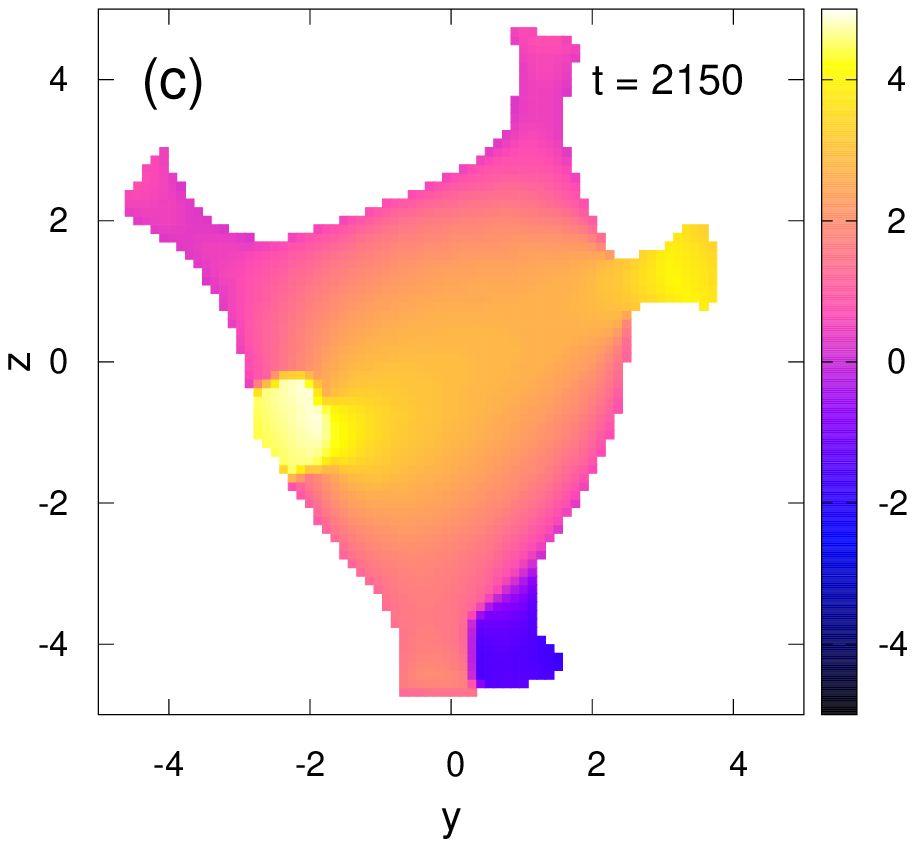, width=75mm}
\end{center}
\caption{Example of a long-term three-dimensional dynamics
of three pairs of attached vortices according to the numerical experiment
with the parameters $a_{11}=88 a_0$, $a_{22}=52 a_0$, and $a_{12}=88 a_0$
characteristic of the ${}^{39}$K-${}^{23}$Na mixture. This gives 
$m=0.6$, $g_{22}=1.0$, and $g_{12}=1.345$. The number of atoms are 
$N_1\approx(l_{\rm tr}/a_{11})\times 326$
and $N_2\approx(l_{\rm tr}/a_{11})\times 107$. 
In contrast to Fig. 1, $\alpha=2.2$ is used.
}
\label{example-3D} 
\end{figure}

\subsection*{Numerical results}

The equations of motion (\ref{GP1}) and (\ref{GP2}) were solved
numerically by the method described in detail in \cite{R2021_attached_v}.
The nonlinear coefficients were taken from real data
on the scattering lengths for 
${}^{39}$K-${}^{87}$Rb  \cite{K39-Rb87-tun},
${}^{41}$K-${}^{87}$Rb  \cite{K41-Rb87-tun-1,K41-Rb87-tun-2}, 
${}^{23}$Na-${}^{39}$K  \cite{Na23-K39-tun}, and
${}^{23}$Na-${}^{87}$Rb \cite{Na23-Rb87-tun} mixtures.

Three pairs of vortex filaments oriented approximately 
along three Cartesian axes exist at the initial
time. This number of vortices was chosen because the
subsequent motion in this case is usually strongly disordered 
and stimulates the separation of vortex pairs.
Since the surface tension confines vortices for a long
time in spite of such conditions, this phenomenon is
indeed worth studying.

In most simulations, the above separation of the
vortex filament from the surface of the core occurred
in a certain time after a quite fast dynamics of vortices.
Immediately before separation, a vortex-antivortex
pair had a V shape, whose base was then separated
from the inner boundary of the shell. The time of this
event $T$ was fixed as the main result and was used as a
rough estimate of the ``quality'' of the chosen combination 
of the parameters. Values $T\gtrsim 1000$, which are
two orders of magnitude larger than the characteristic
period of the trap, can be considered as a good result.
In this time, vortices usually many times change their
mutual positions, constituting long-lived vortex structures. 
Bose--Einstein condensates in real experiments
usually exist in a relatively unchanged form for no more
than several seconds. The dimensionless time $T\approx 630$
at the frequency of the trap about 100 Hz corresponds
to a real time of 1 s.

Figure 1 shows an example of long-term dynamics
in the quasi-two-dimensional regime at the parameters 
$a_{11}=52 a_0$, $a_{22}=99 a_0$, and $a_{12}=83 a_0$
characteristic of the ${}^{23}$Na-${}^{87}$Rb mixture \cite{Na23-Rb87-tun}, 
which gives $m=3.78$, $g_{22}=0.5$, and $g_{12}=1.0$.
The numbers of atoms in this simulation are 
$N_1\approx(l_{\rm tr}/a_{11})\times 204$ and 
$N_2\approx(l_{\rm tr}/a_{11})\times 594$. 
The corresponding video \cite{video1} demonstrates a
complex behavior of vortices, including the formation
of temporal vortex-antivortex pairs, which rapidly
move and collide with other vortices. Partners in pairs
sometimes change in such collisions. Finally, one of
such pairs is separated from the core, as seen in Fig. 1c.
The separation of the vortex filament in a thicker shell
usually occurs earlier as illustrated in the video in \cite{video2}.
It is also noteworthy that close vortices with the same
sign sometimes form temporal pairs, but they cannot annihilate.

Since not only the number of ${}^{23}$Na atoms in the
shell of the ${}^{23}$Na-${}^{87}$Rb system but also the cross 
scattering length can be varied, it is of interest to compare
the corresponding lifetimes. This comparison is given
in Fig. 2. The value $g_{12}=1.0$ is obviously the most
favorable in this case. The lifetime is really a random
value because of the chaotic motion of vortices. Consequently, 
the broken lines in Fig. 2 should be considered 
only as guides for the eye. The main result seen in
this figure is a significant probability of a long lifetime
of a complex vortex system.

Qualitatively similar results were obtained for thin
shells in the ${}^{41}$K-${}^{87}$Rb mixture.

Long lifetimes in significantly three-dimensional
structures were observed, e.g., for ${}^{23}$Na-${}^{39}$K mixtures
\cite{Na23-K39-tun}, where the lighter element ${}^{23}$Na forms the core
and the heavier element ${}^{39}$K constitutes the shell. In
this case, the scattering lengths $a_{\rm K-K}$ and $a_{\rm Na-K}$
strongly depend on the magnetic field (see Fig. 6 in \cite{Na23-K39-tun}). 
A quite large parameter $\alpha=$ 2.1--2.4 ensures
convective stability. The corresponding example is
shown in Fig. 3 and in video \cite{video3}.

The last video \cite{video4} concerns the ${}^{23}$Na-${}^{87}$Rb 
mixture, where the shell consists of ${}^{87}$Rb atoms. Setting
$a_{12}=99 a_0$, we obtain $g_{22}\approx 2.0$ and $g_{12}\approx 2.4$. 
Since $m\approx 0.26$ in this mixture, sufficiently large $\alpha$ values are
necessary for stability. The motion of the vortex configuration
with the parameters $\alpha=7.0$, $n_1=333$, and $n_2=69$ is relatively
regular, as seen in video \cite{video4}.

Large deviations of the shape of the core from the
equilibrium shape are characteristic of three-dimensional 
dynamics, as in the case of equal atomic masses.

Such expressive results have not yet been obtained
for other types of mixtures.

The theoretical consideration with arbitrary
dependences of the scattering lengths on the magnetic
field and with arbitrary ratios of atomic masses reveals
the following approximate symmetry: the three-dimensional 
dynamics of two systems with the same $\alpha$
value but with different $m$ values are qualitatively similar 
(and identically favorable for the storage of vortices) 
if, first, the stratification parameters are identical,
$$
\alpha m^{(1)}/\sqrt{g^{(1)}_{22}}=\alpha m^{(2)}/\sqrt{g^{(2)}_{22}}\approx
\mbox{1.2 --- 1.5},
$$
and, second, the surface tension coefficients are equal,
which can be ensured by the matching of the cross
nonlinear coefficients $g^{(1,2)}_{12}$. Indeed, the hydrodynamic 
parts of the Lagrangians of both systems can be
represented in the same form, and the difference is in
``quantum pressures'' depending on the masses and in
cross nonlinear interactions depending on $g^{(1,2)}_{12}$. 
However, the role of these terms in long-scale dynamics is
reduced only to the surface tension.

This approximate symmetry was confirmed in an
example with the parameters $\alpha=1$,
$m^{(1)}=1$, $g^{(1)}_{22}=0.6$, $g^{(1)}_{12}=1.2$ 
(these values were determined as optimal in \cite{R2021_attached_v})
because long-lived attached vortices were also observed in the system 
with the parameters $m^{(2)}=2$, $g^{(2)}_{22}=2.4$, and $g^{(2)}_{12}=2.4$---$2.6$.

\subsection*{Conclusions}

Thus, numerical examples indicating the possibility 
of observation of long-term dynamics of bubbles
with attached quantum vortices in trapped binary
Bose--Einstein condensates with different atomic
masses have been demonstrated for the first time. The
lifetimes of systems of three pairs of vortices in some
numerical experiments are even longer than those in
the previously simulated ${}^{85}$Rb-${}^{87}$Rb mixture with
approximately equal masses \cite{R2021_attached_v}. 

It is noteworthy that requirements for the properties 
of the core are opposite for the quasi-two-dimensional 
and completely three-dimensional regimes. A hard-trapped, 
almost immobile heavy core over which the less dense shell
easily slips is appropriate for the two-dimensional dynamics of vortices.
A light core with an elastic boundary, which is deformed by the tractive
force induced by a vortex pair, is appropriate for three-dimensional 
dynamics. For the same reason, stratification 
in the three-dimensional case should be not too rigid.

\end{document}